\documentclass[12pt,citerev,final,dvips]{preprint}

\usepackage[final]{epsfig}
\usepackage{rotating}
\usepackage{curves}
\newcommand{\vers}{01.04.96}

\begin{document}
%%%%%%%%%%%%%%%%%%%%%%%%%%%%%%%%%%%%%%%%%%%%%%%%%%%%%%%%%%%%%%%%%
\dateandnumber(March 96){\begin{tabular}[t]{r}
    J{\"u}lich, HLRZ 06/96\\
    Aachen, PITHA 96/11\\
    hep-lat/9604001
    \end{tabular}}
%%%%%%%%%%%%%%%%%%%%%%%%%%%%%%%%%%%%%%%%%%%%%%%%%%%%%%%%%%%%%%%%%
\titleofpreprint%
{Two--dimensional model of}%
{dynamical fermion mass generation}%
{in strongly coupled gauge theories}%
{\version{\em Version 4}}%
{                                                              }%
{                                                              }%
%%%%%%%%%%%%%%%%%%%%%%%%%%%%%%%%%%%%%%%%%%%%%%%%%%%%%%%%%%%%%%%%%
\listofauthors%
{  W.~Franzki$^{1,2,3}$,    }%
{    J.~Jers{\'a}k$^{1,2,4}$, and R.~Welters$^{1,2,5}$          }%
{                                                              }%
%%%%%%%%%%%%%%%%%%%%%%%%%%%%%%%%%%%%%%%%%%%%%%%%%%%%%%%%%%%%%%%%%
\listofaddresses%
{\em $^1$Institute of Theoretical Physics E,
  RWTH Aachen,         D-52056 Aachen, Germany              }%
{\em $^2$HLRZ c/o KFA J{\"u}lich,
      D-52425 J{\"u}lich, Germany                                }%
{
                                                               }%
{}
%%%%%%%%%%%%%%%%%%%%%%%%%%%%%%%%%%%%%%%%%%%%%%%%%%%%%%%%%%%%%%%%%

\abstractofpreprint{%
  We generalize the $N_F=2$ Schwinger model on the lattice by adding a charged
  scalar field. In this so-called $\chi U\phi_2$ model the scalar field
  shields the fermion charge, and a neutral fermion, acquiring mass
  dynamically, is present in the spectrum. We study numerically the mass of
  this fermion at various large fixed values of the gauge coupling by varying
  the effective four-fermion coupling, and find an indication that its scaling
  behavior is the same as that of the fermion mass in the chiral Gross-Neveu
  model.  This suggests that the $\chi U \phi_2$ model is in the same
  universality class as the Gross-Neveu model, and thus renormalizable and
  asymptotic free at arbitrary strong gauge coupling.}

%%%%%%%%%%%%%%%%%%%%%%%%%%%%%%%%%%%%%%%%%%%%%%%%%%%%%%%%%%%%%%%%%
\footnoteoftitle{
%%%%%%%%%%%%%%%%%%%%%%%%%
\footnoteitem($^3$){ \sloppy
E-mail address: wfranzki@hlrz.kfa-juelich.de
}
\footnoteitem($^4$){ \sloppy
E-mail address: jersak@physik.rwth-aachen.de
}
\footnoteitem($^5$){ \sloppy
Present address: AMS, Am Seestern 1, 40547 D\"usseldorf
}%%%%%%%%%%%%%%%%%%%%%%%%%
}
%%%%%%%%%%%%%%%%%%%%%%%%%%%%%%%%%%%%%%%%%%%%%%%%%%%%%%%%%%%%%%%%%
%%%%%%%%%%%%%%%%%%%%%%%%%%%%%%%%%%%%%%%%%%%%%%%%%%%%%%%%%%%%%%%%%
%
%\pagebreak
%\tableofcontents
%
%1111111111111111111111111111111111111111111111111111111111111111111111
%%%%%%%%%%%%%%%%%%%%%%%%%%%%%%%%%%%%%%%%%%%%%%%%%%%%%%%%%%%%%%%%%%%%%%%
%%%%%%%%%%%%%%%%%%%%%%%%%%%%%%%%%%%%%%%%%%%%%%%%%%%%%%%%%%%%%%%%%%%%%%%

\section{Introduction}

Strongly coupled gauge theories tend to break dynamically chiral symmetry, but
fermions which acquire mass through this mechanism are usually confined, as it
is the case in the Schwinger model or in the QCD. From the point of view of
the electroweak symmetry breaking in, or beyond the standard model, a
dynamical mass generation without the fermion confinement is of interest. Such
a situation arises in a class of chiral symmetric strongly coupled gauge
theories on the lattice, in which the gauge charge of the fermion acquiring
mass dynamically is shielded by a scalar field of the same
charge\cite{FrJe95a}.  The question is whether such models are renormalizable
at strong gauge coupling, so that the lattice cutoff can be removed, and the
resulting field theory might be applicable in continuum.

In this work we consider such a lattice model with U(1) chiral symmetry and
vectorlike U(1) gauge symmetry, the $\chi U\phi_d$ model, in $d=2$ dimensions.
It consists of a staggered fermion field $\chi$, a gauge field $U \in$ U(1)
living on the lattice links of length $a$, and a complex scalar field $\phi$
with frozen length $|\phi|=1$. The unconfined fermion field is $F =
\phi^\dagger \chi $. There is no Yukawa coupling, and the mass $am_F$ of the
fermion $F$ arises dynamically. Because of the fermion doubling the fermion
number is $N_F=2$.  The $\chi U\phi_2$ model can be seen either as a
generalization of the Schwinger model with $N_F=2$ by adding a charged scalar
field, or as the 2D scalar QED with added fermions. Similar models have been
studied nearly 20 years ago in the context of the instanton investigations
\cite{CaDa77RaUk78}.

The four dimensional $\chi U\phi_4$ model, considered as a possible theory
with dynamical mass generation of unconfined fermions \cite{FrJe95a}, has been
investigated recently. In spite of some encouraging results
\cite{FrFr95a,FrJe96a} a clarification of its renormalizability properties
remains a difficult task.  Here we demonstrate that in a simpler case, in 2
dimensions, these properties can be investigated with remarkable clarity.
Within the limits of numerical accuracy, we find that the $\chi U\phi_2$ model
is renormalizable at strong gauge coupling $g$, because it belongs to the
universality class of the two-dimensional chiral Gross-Neveu (GN$_2$) model
with $N_F = 2$.

The GN$_2$ model is known to generate fermion mass dynamically at arbitrarily
weak four-fermion coupling $G$, to be nonperturbatively renormalizable, and
asymptotically free. One of its lattice regularizations is identical to the
$\chi U\phi_2$ model in the limit of infinite gauge coupling, $\beta =
1/a^2g^2 = 0$.  In this limit the four-fermion coupling $G$ is an invertible
function $G(\kappa)$ of the hopping parameter $\kappa$ of the scalar field
$\phi$ (or, equivalently, of its bare mass) in the $\chi U\phi_2$ model, with
$G(\kappa) \rightarrow 0$ as $\kappa \rightarrow \infty$.  The scaling
behavior associated with the asymptotic freedom, and the continuum limit of
the GN$_2$ model, are thus obtained as $\kappa \rightarrow \infty$.

These are extremely useful facts when the $\chi U\phi_2$ model is investigated
at a finite gauge coupling, i.\,e., at nonvanishing $\beta$. The idea is to
compare the scaling behavior of the $\chi U\phi_2$ model with that of the
GN$_2$ model, as $\kappa$ grows at fixed $\beta > 0$. For this purpose we
introduce an effective four-fermion coupling $\tilde{G}(\beta,G)$.  This is
now a coupling of the ``composite'' fermions $F = \phi^\dagger \chi$, and thus
characterizes the van der Waals forces arising from the fundamental
interactions between the fields.  At $\beta = 0$ it coincides with the GN$_2$
coupling, $\tilde{G}(0,G) = G$. For $\beta>0$, $\tilde{G}$ is smaller than
$G$, but depends on $\kappa$ similarly as $G(\kappa)$. For the comparison it
is therefore convenient to use $G$ instead of $\kappa$ as one argument of
$\tilde{G}$. We investigate the scaling behavior of $am_F$ with decreasing
$\tilde{G}(\beta,G)$ at various fixed $\beta > 0$, and compare it with the
$\beta = 0$ case.

A determination of $\tilde{G}(\kappa,\beta)$ directly by means of the
four-point function of the composite field would be very expensive. Instead,
we introduce it in an indirect way: we make the assumption that the scaling
behavior of $am_F$ in the $\chi U\phi_2$ model at $\beta > 0$ is described by
the truncated Schwinger-Dyson (SD) equations of the same structure as at
$\beta = 0$ in the GN$_2$ model. The only change is the replacement of $G$ by
$\tilde{G}(G,\beta)$.  Our numerical study of the $\chi U\phi_2$ model at
$\beta > 0$ is mainly concerned with the verification of this assumption.

The use of the SD equations serves further purposes. One is to provide an
analytic framework in which the numerical results obtained at small but
nonzero bare fermion mass $am_0$ can be related to the chiral limit case, $m_0
= 0$, we are actually interested in.  Simulations at $m_0 = 0$ are too time
consuming on larger lattices, and the scaling behavior has to be studied at
$am_0 >0$ by varying both $G$ and $am_0$. The SD equations suggest how to vary
both parameters simultaneously, approaching the critical point, $G = am_0 =
0$, and how to extrapolate to the chiral limit $m_0 = 0$.  We test this
strategy carefully at $\beta = 0$, exploiting the knowledge of the chiral
limit there, and find it remarkably successful. Small deviations, not
accounted for by the SD equations within the used truncation scheme, and
present at very small $am_0$, do not influence the correct scaling behavior
following from the asymptotic freedom.

Further benefit of the use of SD equations is the control of the finite size
effects. Following Ref. \cite{AlGo95}, we solve these equations on lattices of
the same sizes and boundary conditions as those on which numerical data are
obtained. This allows a suitable choice of the data points in the parameter
space, as well as an extrapolation to the infinite volume limit.

Finally, the SD equations allow us to describe the preasymptotic behavior of
the data, obtained for correlation lengths limited by the lattice size, and to
infer the genuine scaling behavior for diverging correlation length.  At
$\beta=0$, we have verified that when the SD equations are used for the
extrapolation, then from those data the correct scaling behavior with $G
\rightarrow 0$ is obtained.

We find that in the $\chi U\phi_2$ model with $\beta \le 1$ the preasymptotic
behavior, and thus presumably also the scaling behavior of the fermion mass
$am_F$ is described quite well by the same SD equations as in the chiral
GN$_2$ model, now with $\tilde{G}(\beta,G)$ replacing $G$. The same scaling
behavior as in the GN$_2$ model is indicated, with $\tilde{G}\rightarrow 0$ as
$\kappa\rightarrow\infty$.

On the basis of this evidence we suggest that the $\chi U\phi_2$ model
belongs, along the critical line at $\kappa = \infty$, to the same
universality class as the GN$_2$ model, and is thus nonperturbatively
renormalizable. The $\chi U\phi_2$ model is therefore an example of a
renormalizable quantum field theory in which the fermion mass is generated
dynamically at strong gauge coupling by the shielded gauge mechanism suggested
in Ref. \cite{FrJe95a}.

In the following section we define the $\chi U \phi_2$ model and explain its
relationship to the GN$_2$ model at $\beta = 0$. In Sec.~3 the phase diagram
is described. The SD equations, and also the determination of the effective
four-fermion coupling $\tilde{G}(\beta,G)$ by means of their inversion, are
discussed in Sec.~4. In the Sec.~5 we test these equations, and the accuracy
of determination of $\tilde{G}$ at $\beta = 0$. In Sec.~6 we then apply the
same method of analysis to the data for $am_F$ at $0 < \beta \le 1$,
demonstrate the applicability of the same SD equations as at $\beta = 0$, and
determine $\tilde{G}(\beta,G)$.  The scaling behavior of the data is
illustrated in Sec.~7. We conclude in Sec.~8 with some remarks about the
meaning of our results.

%22222222222222222222222222222222222222222222222222222222222222222222222
\section{The $\chi U \phi_2$  model and its GN$_2$ limit}

The action of the $\chi U \phi_2$ model consists of three parts:
\begin{equation}
  S_{\chi U \phi} = S_\chi + S_U + S_\phi,
  \lb{CHUPACT}
\end{equation}
where
\begin{eqnarray*}
  S_\chi & \hspace{-2mm} = \hspace{-2mm} & \frac{1}{2} \sum_x \overline{\chi}_x
  \sum_{\mu=1}^2 \eta_{x\mu} (U_{x,\mu} \chi_{x+\mu} - U^\dagger_{x-\mu,\mu}
  \chi_{x-\mu}) %\\ &&
  +am_0 \sum_x \overline{\chi}_x \chi_x \;,\\ 
  S_U & = &\beta \sum_P (1-\mathop{\rm Re}\nolimits {U_P}) \;,\\ 
  S_\phi & = & - \kappa \sum_x \sum_{\mu=1}^2
  (\phi^\dagger_x U_{x,\mu} \phi_{x+\mu} + h.c.) \;.
\end{eqnarray*}
Here, $x$ and $x+\mu$ denote lattice sites and their nearest neighbors,
respectively, and $\chi$ is a staggered fermion field, $\eta_{x\mu}$ being the
usual phase factors.  The link variables $U_{x,\mu} \in$ U(1) represent a
compact abelian gauge field with coupling $\beta=1/a^2g^2$, and $\phi$ is a
complex scalar field with the constraint $|\phi_x|=1$. The fermion and scalar
fields have the same unit charge. The bare fermion mass $m_0$ is introduced
for technical reasons, and we are interested in the limit $m_0=0$, where the
action has a global U(1) chiral symmetry. The phase diagram of this model is
shown in Fig.~\ref{fig:1}.

In the limit case $\beta=0$, one can perform the Lee-Shrock transformation
\cite{LeShr87a}, in which the scalar and gauge fields are integrated out and a
four-fermion term appears.  This leads to the action
\begin{eqnarray}
  \hspace{-5mm} S_{\rm 4f}&=&  - \sum_x \sum_{\mu=1}^2  \Biggl(
  G\, \overline{\chi}_x \chi_x \overline{\chi}_{x+\mu} \chi_{x+\mu}
    - \frac{1}{2}\eta_{x\mu}  \left[ \overline{\chi}_x\chi_{x+\mu}
    - \overline{\chi}_{x+\mu} \chi_x \right] \Biggr) %\nonumber\\&&
  + \frac{am_0}{r} \sum_x
  \overline{\chi}_x\chi_x\;, \lb{GNACT}
\end{eqnarray}
with
\begin{equation}
       G = G(\kappa) = \frac{1-r^2}{4r^2}         \lb{G}
\end{equation}
and
\begin{equation}
       r = \frac{I_1(2\kappa)}{I_0(2\kappa)}\;.  \lb{R}
\end{equation}%
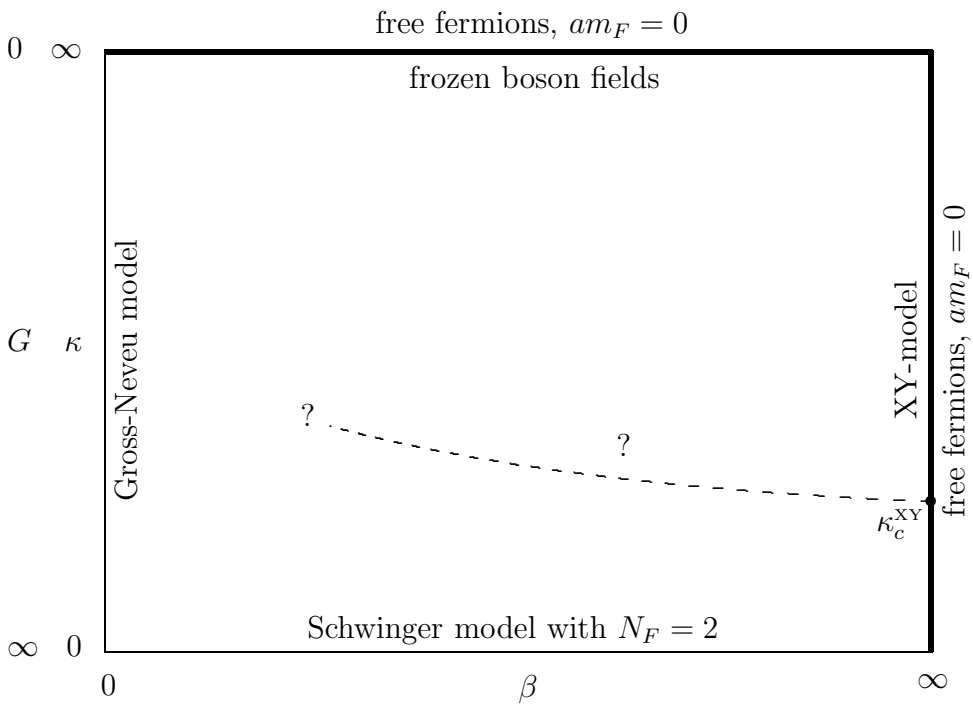
\begin{figure}
\unitlength1cm
\begin{picture}(15,10.3)
%%%%%%%%%%%%%%%%%%%%%%%%% äußerer Rahmen
\thicklines
\put(2,10){\line(1,0){11}}
\put(2,9.97){\line(1,0){11}}
\put(13,2){\line(0,1){8}}
\put(12.97,2){\line(0,1){8}}
\thinlines
\put(2,2){\line(1,0){11}}
\put(2,2){\line(0,1){8}}
%%%%%%%%%%%%%%%%%%%%%%%%% Achsenbeschriftung an kappa-Achse
\put(0.7,1.95){\makebox(1,1)[rb]{$0$}}
\put(0.7,1.95){\makebox(1,1)[lb]{$\infty$}}
\put(0.7,9.9){\makebox(1,1)[rb]{$\infty$}}
\put(0.7,9.9){\makebox(1,1)[lb]{$0$}}
\put(0.7,6){\makebox(1,1)[rb]{$\kappa$}}
\put(0.7,6){\makebox(1,1)[lb]{$G$}}
%%%%%%%%%%%%%%%%%%%%%%%%% Achsenbeschriftung an beta Achse
\put(1.95,0.7){\makebox(1,1)[lt]{$0$}}
\put(12.8,0.7){\makebox(1,1)[lt]{$\infty$}}
\put(7.5,0.7){\makebox(1,1)[lt]{$\beta$}}
%%%%%%%%%%%%%%%%%%%%%%%%% Namen der Spezialfälle an Achsen
\put(4.7,2.1){\makebox(0,0)[lb]{Schwinger model with $N_F=2$}}
\put(5.6,10.15){\makebox(0,0)[lb]{free fermions,  $am_F=0$}}
\put(6.05,9.5){\makebox(0,0)[lb]{frozen boson fields}}
\put( 2.0,4.0){ \begin{sideways} Gross-Neveu model \end{sideways}}
\put(13.0,3.8){ \begin{sideways} free fermions,  $am_F=0$ \end{sideways}}
\put(12.36,5.1){ \begin{sideways} XY-model \end{sideways}}
%%%%%%%%%%%%%%%%%%%%%%%%% PÜ-Linien im Innern
\put(12.985,4){\circle*{0.14}}
\put(12.9,3.9){\makebox(0,0)[rt]{$\kappa_c^{\scriptscriptstyle \rm XY}$}}
\put(4.8,5.3){\makebox(0,0)[rt]{?}}
\put(9.0,4.9){\makebox(0,0)[rt]{?}}
\curvedashes[0.6mm]{0,2,3}
\bezier{2}(13,4)(8,4.2)(5,5)
%\qbezier[40](5,5)(2.5,6.5)(2,10)
%\qbezier[20](5,5)(3.5,5.3)(2,6)
%\qbezier(2,10)(5,9.6)(13,8)
%\qbezier(2,10)(6,9.6)(7,7)
%\qbezier(13,4)(8,4.2)(7,7)
\end{picture}
\vspace{-1.5cm}
\caption{\protect\small
  Phase diagram of the $\chi U \phi_2$ model for $m_0=0$. The well understood
  limit cases are described. The dashed line indicates a possible topological
  phase transition. The fermion mass $am_F$ is nonzero everywhere except on
  the bold-marked boundaries.}
\label{fig:1}
\end{figure}%
\begin{figure}[tbp]
  \begin{center}
    \leavevmode
    \epsfig{file=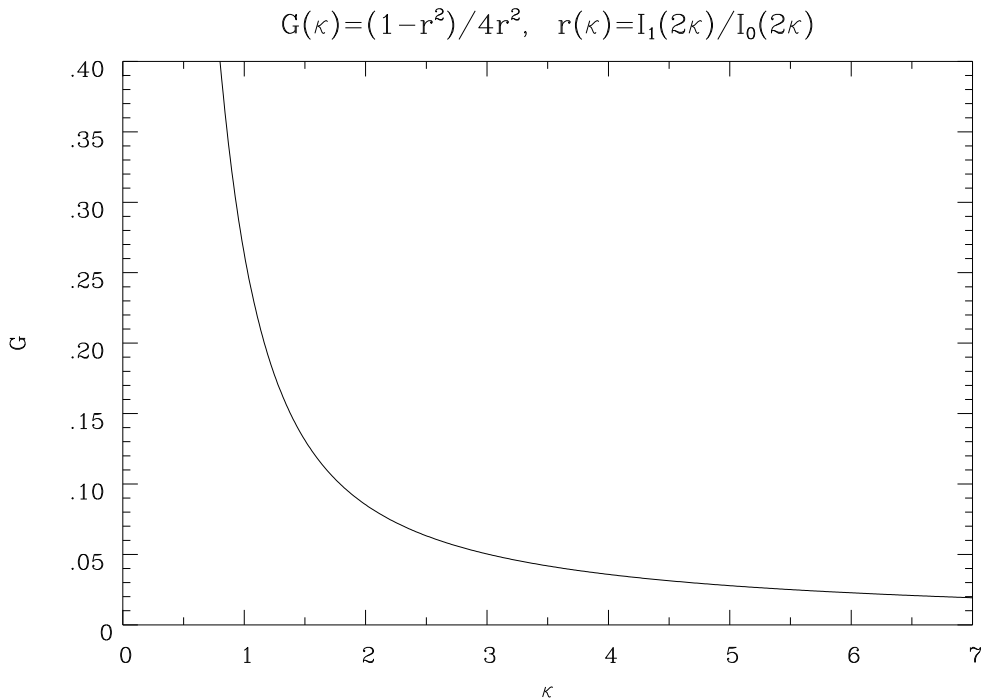,width=10cm,angle=90}
    \vspace{-0.7cm}
    \caption{Relation between the four fermion coupling $G$ and
      $\kappa$ at $\beta=0$.}
    \label{fig:2}
  \end{center}
\end{figure}%
For $m_0=0$ the transformed action (\ref{GNACT}) is that of the chiral
GN${}_2$ model in a certain lattice regularization.  Within the uncertainty of
interpretation of a continuum limit of staggered fermions with a strong gauge
coupling it can be possibly interpreted also as a lattice formulation of the
Thirring model \cite{Ko96}. The GN$_2$ model has a critical point at $G=0$,
where the fermion mass vanishes like
\begin{equation}
  am_F \mathrel{\mathop{\propto}\limits_{G \rightarrow 0}} e^{\displaystyle -
    \frac{\pi}{8G}} \;. \lb{scal}
\end{equation}

The four fermion coupling $G(\kappa)$ is a function of $\kappa$ shown in
Fig.~\ref{fig:2}.  From the point of view of the scaling behavior
(\ref{scal}), the use of $G(\kappa)$ as a parameter instead of $\kappa$ is
very convenient, and we therefore adopt such a reparametrization even at
$\beta > 0$.  There $G$ is not a four-fermion coupling any more, it only
replaces the hopping parameter $\kappa$ according to eq.~(\ref{G}).  Also $r$,
eq.~(\ref{R}), will be understood as a function of $G$ from now on, satisfying
$r(G) \rightarrow 1$ as $G \rightarrow 0$.

We note that the bare mass $am_0/r$ in the action (\ref{GNACT}) of the GN$_2$
model is different from that of the $\chi U\phi_2$ model (\ref{CHUPACT}), as
the field $\chi$ has been rescaled by $\sqrt{r}$ in the course of the
Lee-Shrock transformation.  We use $am_0$ of the $\chi U\phi_2$ model, and the
bare mass in the SD equations for the GN$_2$ model is therefore slightly
$G$-dependent at fixed $am_0$.

We perform the hybrid Monte Carlo simulations on the $V=L^2$ lattices, with
periodic and antiperiodic boundary conditions in first and second $(\mu=t)$
directions, respectively.  The fermion mass $am_F$ is obtained from the gauge
invariant propagator $\langle \phi^\dagger_x \chi_x \phi_y \overline{\chi}_y
\rangle$ by inverting the fermion matrix $M$ and calculating
\begin{equation}
  P_{A B}(k) \Big|_{\vec{k}=0} = 
  \frac{1}{V} \sum_{x,y} e^{\displaystyle i(k+\pi_A)x}
  \phi_x M^{-1}_{xy} \phi^\dagger_y e^{\displaystyle -i(k+\pi_B)y}
  \Bigg|_{\vec{k}=0} \;.
  \lb{fprop1}
\end{equation}
Here $\pi_A$ denotes the usual momentum shifts in the Brillouin zone. The mass
$am_F$, and also the fermion renormalization constant $Z$ result from the fit
in momentum space to
\begin{equation}
  \mathop{\rm Tr}\nolimits\Gamma_t P(k_t)=
  Z \frac{-4i \sin k_t}{\sin^2 k_t + (am_F)^2} \;,
  \lb{fprop2}
\end{equation}
where $\Gamma_t$ is the Golterman-Smit matrix $(\Gamma_{\mu=t})_{AB}$. This
procedure is chiral invariant.

%3333333333333333333333333333333333333333333333333333333333333333333333333333333
\section{Other limit cases and the phase diagram}

The schematic phase diagram for $m_0=0$ is shown in Fig.~\ref{fig:1}.  At
$\kappa = 0$, when the scalar field is decoupled, the $\chi U\phi_2$ model
with $m_0=0$ reduces to the lattice regularized Schwinger model ($d = 2$ QED)
with $N_F = 2$. As is well known, this model is anomaly free, confines
fermions at all $\beta < \infty $, and possesses only massive bosonic states.
Thus $am_F \rightarrow \infty$ as $\kappa \rightarrow 0$. The continuum limit
is expected at the UV fixed point at $\beta = \infty$, and there is no
critical point at $\beta < \infty$.

At $\kappa = \infty$, both the scalar and the gauge fields are
frozen.\footnote{ This is meant in a sense avoiding the Mermin-Wagner-Coleman
  theorem: the limit $\kappa\rightarrow\infty$ is taken before an external
  ``magnetic'' field, required for a definition of the spontaneous symmetry
  breaking, is switched off. We thank G. Roepstorff for a discussion on this
  point.} As is seen in the unitary gauge, the fermion field is decoupled, and
$am_F = am_0$.  At $am_0 = 0$, the line $\kappa = \infty$ is thus a critical
line at which the fermion mass $am_F$ vanishes. The $\beta = 0$ point of this
line is the critical point of the GN$_2$ model.

In the limit $\beta = \infty$, when the gauge coupling vanishes, the fermion
field is decoupled again, and $am_F = am_0$. Thus, in the chiral limit, the
line $\beta = \infty$ is a critical line with vanishing fermion mass, too. The
scalar field variables can be seen as spin variables, and the corresponding
two-dimensional XY$_2$ model is known to have a topological phase transition
at $\kappa = \kappa^{XY}_c \simeq 0.56$.

Let us now consider the inside of the phase diagram in Fig.~\ref{fig:1}. Old
investigations suggest that the model possesses a massive fermion in some
parameter region accessible to the dilute instanton gas \cite{CaDa77RaUk78}.
As follows from the convergence of the strong coupling expansion, at small
nonvanishing $\beta$, the model should have the same properties as at
$\beta=0$. This implies analyticity and nonvanishing $am_F$ for
$\kappa<\infty$ at small $\beta$.

The fate of the topological transition at $\beta=\infty$, as $\beta$ gets
finite, is not completely clear. We have investigated numerically the spectrum
of the model in the vicinity of the dashed line shown in Fig.~\ref{fig:1}.
This line is observable at large $\beta$ as a shallow dip in the masses of the
scalar and vector bosons, which can be constructed from the gauge invariant
products of the type $\phi^\dagger_x U_{x, \mu} \phi_{x+\mu}$. But $am_F$
shows no sensitivity when the dashed line is crossed, and we have found no
state, neither bosonic nor fermionic, indicating a vanishing of the mass in
lattice units on this line at $\beta < \infty$. The dynamical fermion mass
generation at $\beta < \infty$ is thus not influenced by the remnant of the
Kosterlitz-Thoules transition. Presumably, a critical behavior on this
remnant, if any, appears only in some topologically nontrivial observables
\cite{GrIl87KaLa93}. The mass $am_F$ stays finite, and the fermion $F$ gets
infinitely heavy in physical units in any conceivable continuum limit taken on
this line at $\beta < \infty$.

We have checked that there is no indication of any other phase transition, not
even of some change of behavior of some local observable, anywhere else in the
phase diagram of the $\chi U\phi_2$ model.  Our data thus indicate that the
fermion mass $am_F$ is nonzero for any finite $\beta$ and $\kappa$. It
decreases when any of these parameters increases and one of the boundaries
bold-marked in Fig.~\ref{fig:1} is approached. As in the GN$_2$ model, the
fermion mass generation takes place without the spontaneous chiral symmetry
breaking, which is forbidden by the Mermin-Wagner-Coleman Theorem. A continuum
limit taken on the line $\beta = \infty$ might lead to an interesting
generalization of the Schwinger model in the continuum. In this work, we
concentrate on the scaling behavior of $am_F$ when the line $\kappa = \infty$
is approached.

%4444444444444444444444444444444444444444444444444444444444444444444444444444
\section{Schwinger-Dyson equations for the GN${}_2$ model}

The SD equations for the fermion propagator in a four-fermion theory,
truncated after O($G$), can be represented graphically as
\begin{equation}
\unitlength1cm
\begin{picture}(15,4)
%%%%%%%%%%%%%%%% voller Propagator =
\put(0,2){\vector(1,0){2.5}} \put(2.5,2){\line(1,0){0.5}}
\put(1.5,2){\circle*{1}}
\put(3.35,1.9){\makebox(0.3,0.3)[bl]{=}}
%%%%%%%%%%%%%%%% nackter Propagator +
\put(4,2){\vector(1,0){1.5}} \put(5.5,2){\line(1,0){1.5}}
\put(7.35,1.84){\makebox(0.3,0.3)[bl]{+}}
%%%%%%%%%%%%%%%% Term (a) +
\put(8,2){\vector(1,0){0.5}} \put(8.5,2){\line(1,0){0.45}}
\put(9.05,2){\vector(1,0){1.75}} \put(10.8,2){\line(1,0){0.2}}
\put(10.0,2){\circle*{1}}
\put(8.95,2){\line(0,1){0.5}} 
\put(9.05,2){\line(0,1){0.2}} 
\put(9.05,2.5){\line(0,-1){0.3}} %\put(9.05,2.5){\vector(0,-1){0.3}}
\put(8.95,2.75){\oval(0.5,0.5)[l]} \put(9.05,2.75){\oval(0.5,0.5)[r]}
\put(9,3){\circle*{1}}
\put(9.3,2.68){\vector(0,-1){0.01}}
\put(11.35,1.84){\makebox(0.3,0.3)[bl]{+}}
%%%%%%%%%%%%%%%% Term (b)
\put(12,2){\vector(1,0){0.5}} 
\put(12.5,2){\vector(1,0){2.3}} \put(14.8,2){\line(1,0){0.2}}
\put(14,2){\circle*{1}}
\put(12.7,2.1){\line(1,0){0.3}} 
\put(13.3,2.1){\line(-1,0){0.3}} %\put(13.3,2.1){\vector(-1,0){0.3}}
\put(12.7,2.35){\oval(0.5,0.5)[l]} \put(13.3,2.35){\oval(0.5,0.5)[r]}
\put(12.7,2.6){\line(1,0){0.6}}
\put(13,2.6){\circle*{1}}
\put(13.55,2.28){\vector(0,-1){0.01}}
\put(9.2,1){\makebox(0.5,0.3)[lb]{($a$)}}
\put(13.2,1){\makebox(0.5,0.3)[lb]{($b$)}}
\end{picture}\;\raisebox{18mm}{.}
  \lb{figsd}
\end{equation}
On a finite lattice they read (see e.\,g.~\cite{AlGo95}):
\begin{equation}
N=\frac{am_0}{r(G)}+ \frac{4G}{V} \sum_k
 \frac{N}{\sum_\nu F_\nu^2 \left( \sin(k_\nu a)
   \right)^2 + N^2}\;, \lb{SDN}
\end{equation}
\begin{equation}
F_\mu = 1+\frac{2G}{V} \sum_k \frac{F_\mu \left(
    \sin(k_\mu a)  \right)^2 } {\sum_\nu  F_\nu^2  \left(
    \sin(k_\nu a)  \right)^2 + N^2}\;.  \lb{SDF}
\end{equation}
These are three coupled equations for  $N$ and $F_\mu$
which we solve numerically. Then
\begin{equation}
am_F=N/F_t\;, \lb{MF}
\end{equation}
\begin{equation}
Z = \frac{1}{rF_t} \lb{Z}
\end{equation}
are determined.
If the term (b) in (\ref{figsd}) is neglected, one obtains the gap equation
with $F_\mu=1$ and $am_F=N$.

In the infinite volume, when $F_1=F_2=F$, and for small $G$, the approximate
analytic solution of the SD equations is
\begin{eqnarray}
  N &=& \frac{am_0}{r} -\frac{8G N}{\pi F^2}
  \ln \left( \frac{2N}{\pi F} \right)\;,
\lb{eq:N} \\
F &=& \frac{1}{2} + \frac{1}{2} \sqrt{1+G} \;. \lb{eq:F}
\end{eqnarray}
For $m_0=0$, one obtains the scaling behavior
(\ref{scal}) as $G \rightarrow 0$.

The idea of the combined limit $G \rightarrow 0$ and $am_0 \rightarrow 0$ is
to make $am_0/r$ a function of $G$ in such a way that eq.\ (\ref{eq:N}) is
solvable, and that $am_0(G) \rightarrow 0$ as $G\rightarrow 0$.  We choose
\begin{equation}
  \frac{am_0(G,s)}{r(G)} = (1-s)\frac{\pi F}{2}e^{\displaystyle
    -\frac{\pi F^2 s}{8 G}}\;,     \lb{MBARE}
\end{equation}
using the approximate solution (\ref{eq:F}) for F. Here $s$ is a free
parameter obeying $0 < s \le 1$. For this choice of $am_0$ the solution of
eq.~(\ref{eq:N}) is
\begin{equation}
  N=\frac{\pi F}{2} e^{\displaystyle -\frac{\pi F^2 s}{8 G}}\;,
\end{equation}
and thus
\begin{equation}
  am_F=\frac{N}{F}=\frac{\pi}{2} e^{\displaystyle -\frac{\pi F^2 s}{8 G}}.
   \lb{SOLU}
\end{equation}
The asymptotic scaling law along the lines of constant $s$ is
\begin{eqnarray}
  am_F &=& \frac{\pi}{2}e^{\displaystyle -\frac{\pi s}{8 G}},\lb{SCALING}\\
  \frac{am_F}{am_0} &=& \frac{1}{1-s} \;,  \lb{RATIO}
\end{eqnarray}
where we have used $rF\rightarrow 1$ as $G\rightarrow 0$.  Obviously, the
ratio $m_0/m_F$ does not vanish except if $s=1$. For $s<1$, a continuum GN$_2$
model with nonvanishing bare mass is obtained. We therefore distinguish
between the chiral limit, $m_0=0$ for any $a$, and the statement $am_0=0$,
which at a critical point with $am_F=0$ allows $a=0$, $m_0/m_F \ne 0$.

It may appear strange that the scaling behavior (\ref{SCALING}) depends on
$s$, i.\,e.\ on the path, though the same critical point is approached.  But
this is similar to the scaling behavior in a magnetic system described by the
equation of state
\begin{equation}
  h=M^\delta f\left( \frac{t}{M^{1/\beta}} \right),
  \lb{widom1}
\end{equation}
when the external field $h$ (corresponding to $am_0/r$) and the reduced
temperature $t$ (corresponding to $G$) are varied simultaneously. For example,
if
\begin{equation}
  h = c \cdot t^p\;,\hspace{1cm} p \le \beta\delta\;,
  \lb{widom4}
\end{equation}
the scaling law is $p$-dependent,
\begin{equation}
  M \propto t^{p/\delta}\;.
  \lb{widom5}
\end{equation}
In our case the scaling is described by an essential singularity instead of
the power law (\ref{widom5}).  Needless to say, the SD equations reproduce
correctly the first coefficient of the $\beta$-function of the GN$_2$ model
with nonvanishing bare mass since the one loop contribution is taken into
account correctly.

The truncated SD equations cannot be expected to describe the data correctly
at very small $am_0$, because the contributions of the neglected fermion loops
increase with decreasing $am_0$. Then only the exponential scaling behavior,
eq.~(\ref{SCALING}), but not the value of the constant prefactor in
(\ref{SCALING}), and thus of the ratio (\ref{RATIO}), are predicted correctly.
We shall take this discrepancy into account by relying in our determination of
$\tilde{G}$ in Sec.~\ref{effcoup} on the results at larger values of $am_0$
(mostly $am_0=0.4$).

An important step in our data analysis at $\beta > 0$ is the use of the
inverse SD equations. For a chosen $\beta$, $s$ and $am_0$, the fermion mass
$am_F$, obtained on a certain lattice, is inserted into the corresponding SD
equations (\ref{SDN}), (\ref{SDF}) by using eq.~(\ref{MF}).  The four fermion
coupling is considered as a free parameter $\Gamma$. It is determined by
solving the three equations (\ref{SDN}), (\ref{SDF}), taken as equations
determining $\Gamma$ and $F_\mu$:
\begin{equation}
am_F F_t=\frac{am_0}{r(\Gamma)}+\frac{4\Gamma}{V} \sum_k
 \frac{am_F F_t}{\sum_\nu F_\nu^2 \left( \sin(k_\nu a)
   \right)^2 + (am_F F_t)^2}\;, \lb{ISDN}
\end{equation}
\begin{equation}
F_\mu = 1+\frac{2\Gamma}{V} \sum_k \frac{F_\mu \left(
    \sin(k_\mu a)  \right)^2 } {\sum_\nu  F_\nu^2  \left(
    \sin(k_\nu a)  \right)^2 + (am_F F_t)^2}\;.  \lb{ISDF}
\end{equation}
The resulting values of $\Gamma$ can, in principle, depend on all the
parameters, $\Gamma = \Gamma(\beta,G,am_0,V)$.

%5555555555555555555555555555555555555555555555555555555555555555555555555
\section{Test of the SD equations in the GN$_2$ limit}

The aim of this section is to investigate how the known scaling behavior
(\ref{scal}) of the GN$_2$ model can be confirmed by the data for $am_F$
obtained in numerical simulation of the $\chi U\phi_2$ model at $\beta = 0$.
The fermion matrix inversion has turned out to be very slow at $m_0 = 0$ on
large lattices ($\ge 32^2$), and the simulations have to be performed at
finite $am_0$. We have made several attempts to extrapolate to $m_0 = 0$ the
values of $am_F$ obtained for several $am_0$ at fixed $G$. However, using
e.\,g.\ a power law we failed to reproduce reliably the values obtained by
long simulations directly at $m_0 = 0$.

Therefore we have adopted the strategy of the combined approach to the
critical point $G = am_0 = 0$, in which $G$ and $am_0$ vary simultaneously,
choosing the paths $s = const$, eq.~(\ref{MBARE}), suggested by the SD
equations. The chosen values of $s$ are $s = 0.2, 0.3, ..., 0.7$. The
simulations have been performed at the values of $am_0$ and $G$ satisfying the
relation (\ref{MBARE}), and chosen such that the value of $am_F$ predicted by
the SD equations satisfies
\begin{equation}
             \frac{1}{am_F} < \frac{L}{4}\;.   \lb{L4} 
\end{equation}
This restriction turned out to be necessary in order to avoid a finite size
dependence both of the predicted and measured values of $am_F$.

\begin{figure}[tbp]
 \begin{center}
  \leavevmode
%  \hspace*{-0.7cm}
    \epsfig{file=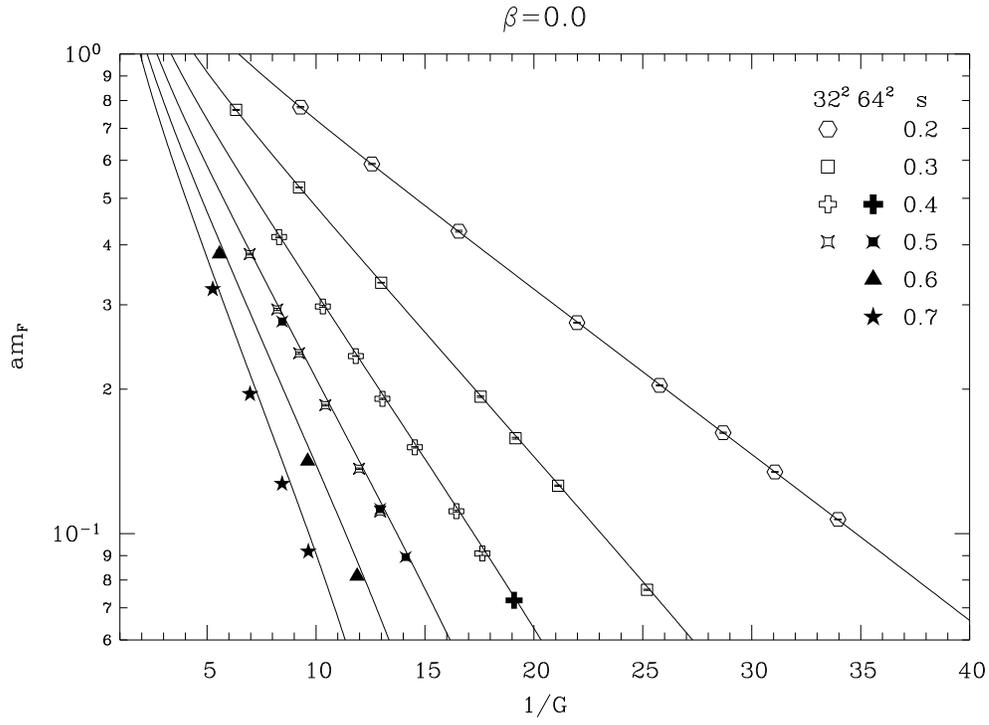,width=10cm,angle=90}
%    \vspace{-1.0cm}
    \caption{The data for $am_F$ at $\beta=0$ for various fixed $s$ compared
      with the predictions of the SD equations.}
    \label{fig:3}
 \end{center}
\end{figure}

In Fig.~\ref{fig:3} we show the semi-logarithmic plot of the data for $am_F$
against $1/G$, and compare them with the prediction of the SD equations
(\ref{SDN}) and (\ref{SDF}) on the $32^2$ and $64^2$ lattices. It demonstrates
that the SD equations describe the data for smaller $s$ very well, and for
larger $s$, $s = 0.6, 0.7$, still quite well, though for these $s$ the bare
mass is very small. But one cannot see that the data is still far from the
asymptotic scaling behavior (\ref{SCALING}), and that the seemingly linear
decrease does not have the right slope. Also the size of the error bars is
barely visible on the logarithmic scale.

\begin{figure}[tbp]
 \begin{center}
  \leavevmode
%  \hspace*{-0.7cm}
    \epsfig{file=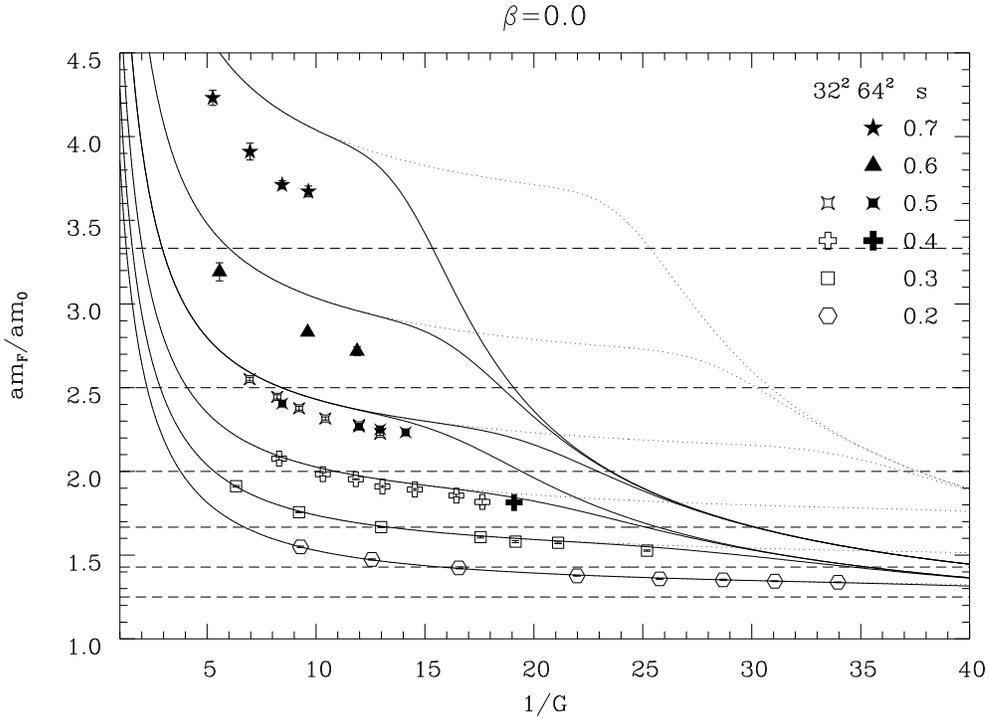,width=10cm,angle=90}
%    \vspace{-1.0cm}
    \caption{The ratio $am_F/am_0$ at $\beta=0$ for various fixed $s$ compared
      with the SD equations on $32^2$ ($s\leq 0.5$) and $64^2$ ($s\geq 0.5$)
      lattices (full lines), and $1024^2$ lattice (dotted lines). The dashed
      horizontal lines represent the expected asymptotic behavior.}
    \label{fig:4}
 \end{center}
\end{figure}%
Therefore, in Fig.~\ref{fig:4} we show the data as ratios $am_F/am_0$ plotted
on the linear scale. The curves are the SD predictions for this ratio on the
$32^2$ and $64^2$ lattices (full lines) and $1024^2$ lattice (dotted lines),
whereas the dashed horizontal lines represent the expected asymptotic scaling
behavior, eq.~(\ref{RATIO}). The benefits of the SD equations become manifest:
For smaller $s$ the agreement with the data is within the tiny but now visible
error bars.  With increasing $s$ the agreement gets worse, as expected for
decreasing $am_0$. Nevertheless, the SD equations still reproduce the
$1/G$-dependence qualitatively. They predict the onset of finite size effects,
manifested by a downward bend of the curves. These effects set on at larger
$1/G$ on larger lattices. We have checked that the data not satisfying the
restriction (\ref{L4}) behave in a similar way. Furthermore, one can observe
the now apparent difference from the asymptotic scaling behavior
(\ref{RATIO}), with an indication how this behavior would be slowly achieved
on huge lattices (dotted lines). We find that the SD equations describe the
data for $s\leq 0.5$ long before the asymptotic scaling sets on. This allows
us to extrapolate the data obtained on achievable lattices to large $1/G$, and
to interpret them as an evidence for the asymptotic scaling
(\ref{SCALING})--(\ref{RATIO}), expected for the GN$_2$ model with nonzero
bare mass.

\begin{figure}[tbp]
 \begin{center}
   \leavevmode
%  \hspace*{-0.7cm}
   \epsfig{file=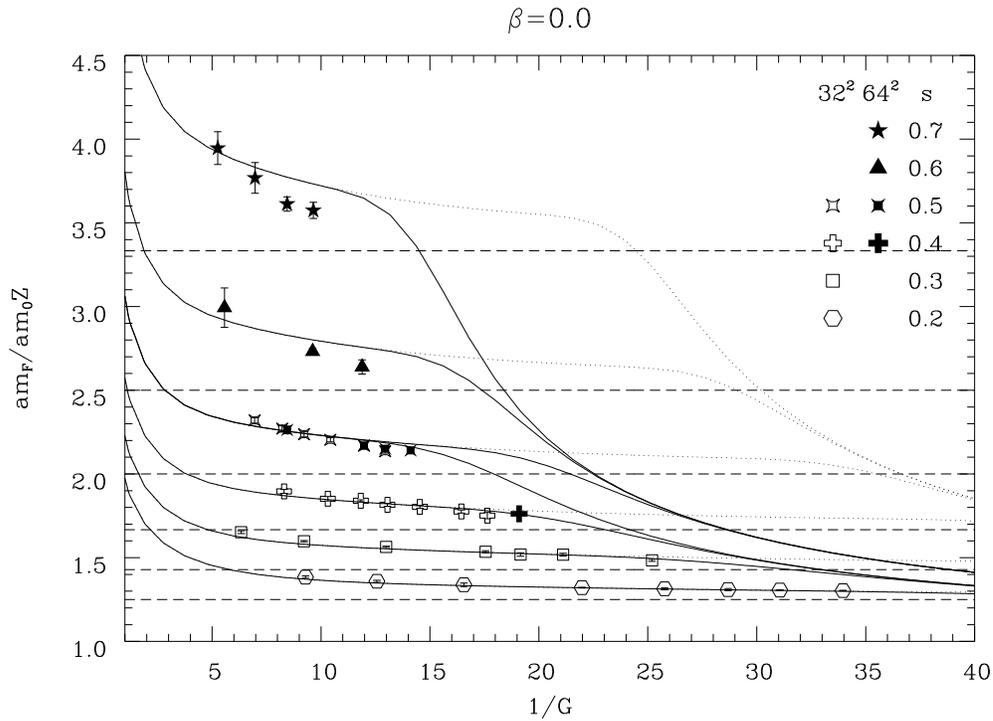,
     width=10cm,angle=90}
%    \vspace{-1.0cm}
   \caption{As in Fig.~\protect\ref{fig:4}, but for the ratio $am_F/am_0Z$,
     $Z$ obtained from the fermion propagators (data) and from
     (\protect\ref{Z}) (curves).}
   \label{fig:5}
 \end{center}
\end{figure}

We expect that the discrepancies observed at $s=0.6$ and 0.7 are due to the
truncation, and thus do not indicate any deviation from the asymptotic
scaling.  This is further supported by the observation that the agreement
between the data and the SD equations is improved even at larger $s$, if we
plot the data for $am_F/am_0Z$, with $Z$ measured by means of the relation
(\ref{fprop2}), and compare them with the SD equation using (\ref{Z}) (see
Fig.~\ref{fig:5}).  Because the measured values of Z are consistent with $Z
\rightarrow 1$ for $G \rightarrow 0$, as follows also from eq.~(\ref{Z}), the
asymptotic behavior is not changed, and we now see how it is approached even
for larger~$s$.

\begin{figure}[tbp]
 \begin{center}
  \leavevmode
%  \hspace*{-0.7cm}
    \epsfig{file=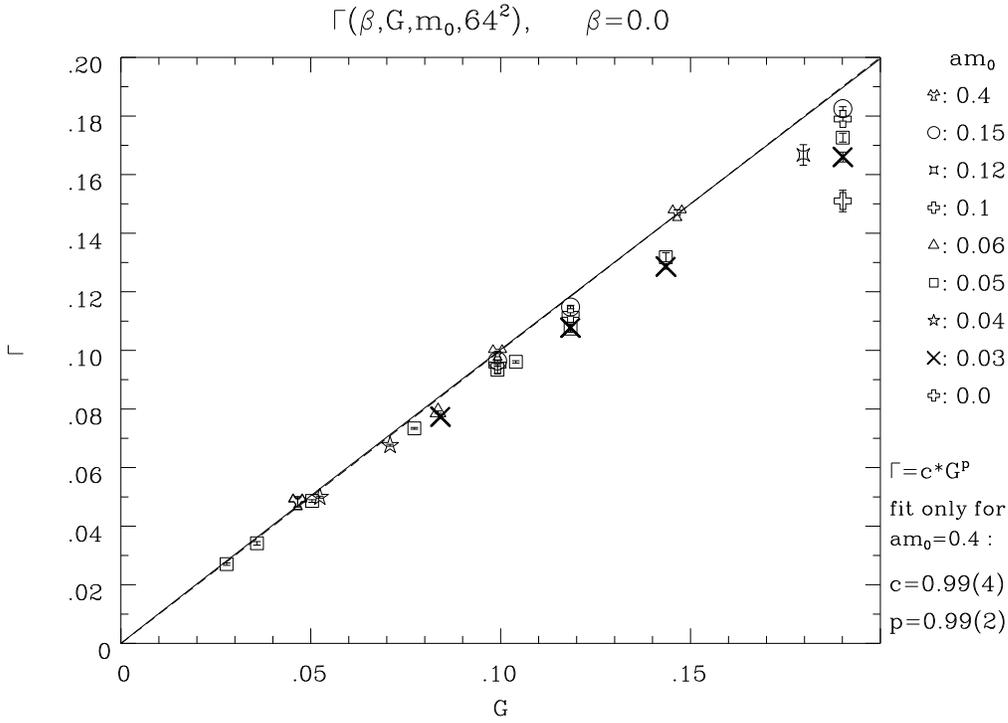,width=10cm,angle=90}
%    \vspace{-1.0cm}
    \caption{The values of $\Gamma(\beta,G,am_0,V)$ obtained by the inversion
      of the SD equations at $\beta=0$ for $V=64^2$. The full line
      representing the fit by means of eq.~(\ref{GG}) to the $am_0=0.4$ data
      nearly coincides with the diagonal (dashed) line.}
    \label{fig:6}
 \end{center}
\end{figure}

As a preparation for the studies at $\beta >0$, it is illustrative to test, at
$\beta = 0$, also the inversion of the SD equations according to
eqs.~(\ref{ISDF}) and (\ref{ISDN}).  The above data for $am_F$ has been used
as input to these equations, and $\Gamma(0,G,am_0,V)$ has been obtained by
their numerical solution. The results are shown in Fig.~\ref{fig:6}, where
they are compared with the expected value of the coupling $\tilde{G}(0,G) =
G$. We observe that for a rather large bare mass, $am_0 = 0.4$, the agreement
is excellent. The fit $\Gamma(0,G,0.4,V) =c G^p$ gives, for $V = 32^2$ and
$64^2$, results consistent with $c = p = 1$ (see table~1).

For smaller $am_0$, deviations of $\Gamma$ from the true coupling $G$ are
observable.  Of course, these deviations are of the same origin as those in
Figs.~\ref{fig:3} and \ref{fig:4}: they reflect the inaccuracy of the SD
equations for larger $s$, and thus for smaller $am_0$. For a quantitative
comparison with the SD equations it is therefore advantageous to perform
simulations at not too small $am_0$. Nevertheless, even for smaller $am_0$ the
deviations remain small. The degree of agreement between $\Gamma$ and $G$ at
$\beta=0$ can serve at $\beta > 0$ as an estimate for the accuracy, with which
one can determine the effective coupling $\tilde{G}(\beta,G)$ from the values
of $\Gamma(\beta,G,am_0,V)$ obtained by the inversion of the SD equations.

%6666666666666666666666666666666666666666666666666666666666666666666
\section{Effective four-fermion coupling at $\beta > 0$}
\label{effcoup}
The experience gained at $\beta = 0$ makes it clear that also at $\beta > 0$
the simulations have to be performed at $am_0 >0$, and that on lattices of
affordable sizes there is no chance to observe the asymptotic scaling
directly. Therefore, we investigate the scaling behavior of the fermion mass
$am_F$ at $\beta > 0$ by means of the following strategy:

\begin{enumerate}
\item The conjecture that the $\chi U\phi_2$ model at $\beta > 0$ belongs to
  the same universality class as the GN$_2$ model suggests the use of the same
  SD equations as at $\beta = 0$ for the description of the nonasymptotic data
  for $am_F$. The only foreseen difference is the value of the effective four
  fermion coupling $\tilde{G}(\beta,G)$, which is not any more a known
  function (\ref{G}) of $\kappa$, but has to be determined from the data. A
  similar idea was very successful in the earlier study of the universality
  class of the two-dimensional Yukawa model \cite{DeFo93bFoFr94}.
  
\item For each value of $am_F$, obtained at some $\beta$, $G$ and $am_0$ on a
  lattice of volume $V$, we invert the SD equations (\ref{ISDN}) and
  (\ref{ISDF}), obtaining $\Gamma(\beta,G,am_0,V)$.
  
\item Of course, in principle the effective coupling should be independent of
  $am_0$ and $V$. Because of the limited accuracy of the truncated SD
  equations, some dependence of $\Gamma(\beta,G,am_0,V)$ on $am_0$ remains,
  however. On the basis of the observation that at $\beta = 0$ the values of
  $\Gamma$ and $G$ are consistent for $am_0 = 0.4$, we assume that also at
  $\beta > 0$ the values of $\tilde{G}(\beta,G)$ can be obtained for this
  $am_0$, and define
  \begin{equation}
    \tilde{G}(\beta,G) = \Gamma(\beta,G,am_0=0.4,V).  \lb{GGAM}
  \end{equation}
  This effective four-fermion coupling is $V$-independent, provided $V$ is
  sufficiently large.
  
\item At each $\beta$, we determine the $G$-dependence of
  $\tilde{G}(\beta,G)$, which turns out to be consistent with a power law
  \begin{equation}  
    \tilde{G}(\beta,G) = c(\beta) G^{p(\beta)}.  \lb{GG}
  \end{equation}
  
\item If $\tilde{G}(\beta,G)$ were known prior to the numerical calculations
  of $am_F$ at $\beta > 0$, one would choose the data points so that in the
  $(am_0,G)$ plane they lie on lines similar to (\ref{MBARE}), with $G$
  replaced by $\tilde{G}(\beta,G)$. The same simple comparison with the
  expected scaling behavior (\ref{SCALING}), now with $\tilde{G}(\beta,G)$,
  would then be possible. Without this knowledge, we have decided to acquire
  the data on the same lines of constant $s$, eq.~(\ref{MBARE}), as at $\beta
  = 0$. As will be explained in the next section, one can then recalculate the
  predictions of the SD equations with $\tilde{G}(\beta,G)$ to the lines
  $s=\mbox{const.}$, and observe the approach to the scaling behavior, though
  these lines are now less suitable for this purpose.

\end{enumerate}

We have determied $am_F$ at $\beta = 0.2, 0.3, 0.5, 0.7, 1.0$ on the $32^2$
lattice for $s=0.2,$ 0.3, 0.4, 0.5, and at $\beta = 0.3$ also on the $64^2$
lattice for $s=0.4, 0.5, 0.6, 0.7$. The intervals of the $G$ and $am_0$ values
have been chosen such that the fermion mass is consistent with the requirement
(\ref{L4}).  As expected, the fermion mass $am_F$ decreases with increasing
$\beta$ at constant $G$, because of the decreasing gauge coupling.

For $\beta > 1$ the mass $am_F$ is measurable only in a small $G$-interval
below or in the vicinity of the dashed line in Fig.~\ref{fig:1}. Nevertheless,
one can see that $am_F$ is insensitive to this line.

\begin{figure}[tbp]
  \begin{center}
    \leavevmode
    %% \hspace*{-0.7cm}
    \epsfig{file=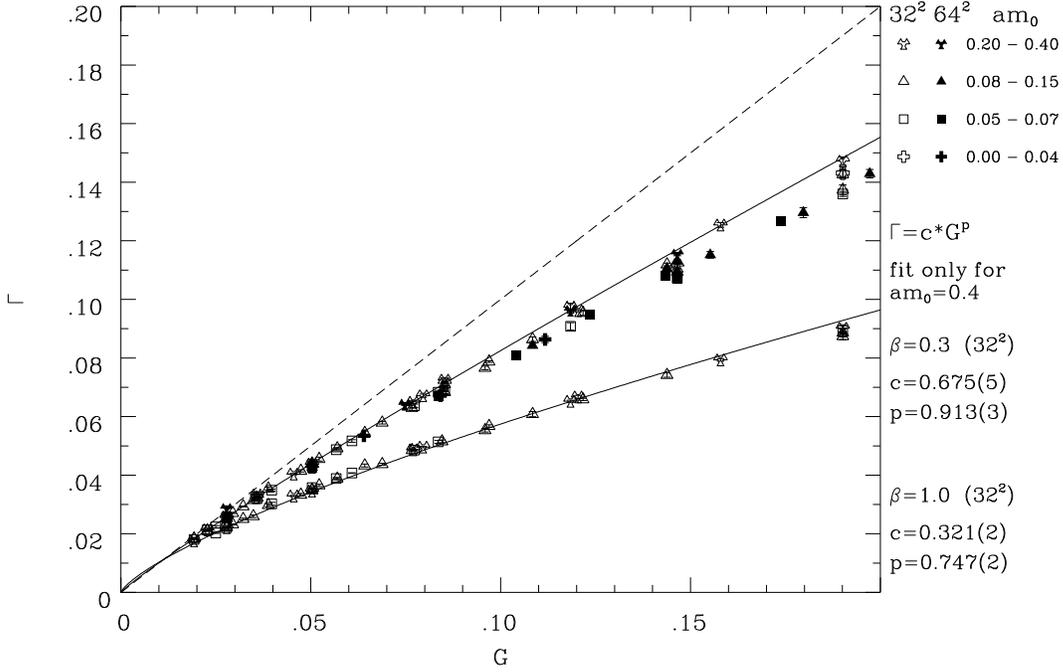,
      width=10cm,angle=90}
    %% \vspace{-1.0cm}
    \caption{The values of $\Gamma(\beta,G,am_0,V)$ at $\beta=0.3$ and
      $\beta=1.0$. The full lines are fits by means of eq.~(\ref{GG}) to the
      $am_0=0.4$ data. They represent the obtained effective coupling
      $\tilde{G}(\beta,G)$.}
    \label{fig:7}
  \end{center}
\end{figure}

In Fig.~\ref{fig:7} we show the results for $\Gamma(\beta,G,am_0,V)$ at $\beta
= 0.3$ and $\beta = 1$. They cluster for each $\beta$ along, or slightly below
a simple curve. This curve is the power law fit (\ref{GG}) to the values of
$\tilde{G}(\beta,G)$, determined according to eq.~(\ref{GGAM}) at $am_0 =
0.4$. The results for smaller $am_0$ do not deviate from this curve more than
in the $\beta = 0$ case (Fig.~\ref{fig:6}). For other $\beta$ values the
results are very similar.

Fits by means of the power law (\ref{GG}) describe the data at $am_0 = 0.4$
very well. In table~\ref{tab:1} we present the results for the fit parameters
$c(\beta)$ and $p(\beta)$.
\begin{table}[tbp]
 \begin{center}
$
\begin{array}{|r|r|r|r|r|}
  \hline
   \multicolumn{1}{|c|}{}
   & \multicolumn{2}{c|}{32^2} & \multicolumn{2}{c|}{64^2}  \\
   \cline{2-5}
   \multicolumn{1}{|c|}{\mbox{\raisebox{1.2ex}[1.2ex]{$\beta$}}} &
   \multicolumn{1}{c|}{c} &
    \multicolumn{1}{c|}{p}   & \multicolumn{1}{c|}{c} &
    \multicolumn{1}{c|}{p}      \\ \hline
    0.0 & 1.038(7)   & 1.016(4)   & 0.99(4)    & 0.99(2)    \\ \hline
    0.2 & 0.767(9)   & 0.940(5)   &            &            \\ \hline
    0.3 & 0.675(5)   & 0.913(3)   & 0.62(1)    & 0.88(1)    \\ \hline
    0.5 & 0.512(6)   & 0.844(4)   &            &            \\ \hline
    0.7 & 0.422(3)   & 0.806(2)   &            &            \\ \hline
    1.0 & 0.321(2)   & 0.747(2)   &            &            \\ \hline
\end{array}
$
\caption{The values of the parameters determining
  the effective coupling $\tilde{G}(\beta,G)$ by means of eq.~(\ref{GG}). See
  the text for a discussion of the errors.}
\label{tab:1}
\end{center}
\end{table}
The indicated errors come from the Minuit fit for $\tilde{G}(\beta,G)$ by
means of eq.~(\ref{GG}). As the values of $\tilde{G}(\beta,G)$ result from a
data analysis in several steps, these errors are too naive. A more realistic
error estimate might be the difference between the results on $32^2$ and
$64^2$ lattices at $\beta=0$, where we known that these results should be
consistent. This suggests the errors of sizes $\Delta c/c\simeq 0.05$ and
$\Delta p/p\simeq 0.03$.

These results demonstrate that the introduction of the effective four fermion
coupling $\tilde{G}(\beta,G)$ by means of the SD equations is sensible.  In
other words, one can find such a function $\tilde{G}(\beta,G)$, that the data
is consistent with the SD equations when $\tilde{G}(\beta.G)$ is used as a
coupling. Assuming now that this is true also at smaller $G$ and $am_0$,
i.\,e.\ beyond the intervals we could investigate, we can deduce the scaling
behavior of $am_F$ with $\tilde{G}(\beta,G)$ from these SD equations.

The power law dependence (\ref{GG}) of $\tilde{G}(\beta,G)$ on $G$ is
questionable for very small $G$ because of the singular behavior at $G=0$, and
probably should not be trusted there. K.-I.  Kondo has investigated in the
continuum a model quite similar to the $\chi U\phi_2$ model, both in $d=4$
\cite{Ko96} and $d=2$ \cite{Koxx}. Recently, he has obtained in $d=2$, by
solving the SD equations for the full $\chi U\phi_2$ model, the effective
four-fermion coupling $\tilde{G}$ as a series in $G$ \cite{Koxx}. This would
contradict (\ref{GG}).  We have checked that our results for $\tilde{G}$ can
be described by a polynomial in $G$, too, but the coefficients of the higher
powers of $G$ are large and unstable, and such an analytic description is
therefore of little use.

%7777777777777777777777777777777777777777777777777777777777777777777
\section{Scaling behavior at $\beta > 0$}

According to the results presented in the previous section, the SD equations
with $G$ replaced by $\tilde{G}(\beta,G)$ describe the data for $am_F$ for
larger $am_0$ values. These equations predict also the behavior of $am_F/am_0$
as $\tilde{G}$ and $am_0$ approach zero along the lines
\begin{equation}
  \frac{1}{r(\tilde{G})}am_0(\tilde{G},\tilde{s}) 
      = (1-\tilde{s})\frac{\pi F}{2}e^{\displaystyle
        -\frac{\pi F^2 \tilde{s}}{8 \tilde{G}}}\;,\;\;\tilde
      s=\mbox{const.}\;,
      \lb{MBARET}
\end{equation}
in the ($am_0$,$\tilde{G}$) plane. Along these lines the asymptotic scaling
behavior is analogous to (\ref{SCALING}),
\begin{equation}
  am_F = \frac{\pi}{2}e^{\displaystyle -\frac{\pi \tilde{s}}{8
      \tilde{G}}}\;,\hspace{1cm} \frac{am_F}{am_0}= \frac{1}{1-\tilde{s}} \;.
  \lb{SCALINGT}
\end{equation}
This is the same scaling behavior as in the GN$_2$ model with $am_0 \ge 0$,
when $G$ is replaced by $\tilde{G}$. From eqs.~(\ref{G}), (\ref{R}) and
(\ref{GG}) it follows that $\tilde{G}(\beta,G) \rightarrow 0$ as $\kappa
\rightarrow \infty$.  The successful analysis of the data for $am_F$ at $\beta
> 0$ by means of the SD equations with $\tilde{G}(\beta,G)$ is thus an
indication that the $\chi U\phi_2$ model at $\beta > 0$ belongs to the same
universality class as the GN$_2$ model, when the critical line $\kappa =
\infty$ is approached.  In the chiral limit, the scaling behavior predicted by
the SD equations with $\tilde{G}(\beta,G)$ is
\begin{equation}
  am_F \propto e^{\displaystyle -\frac{\pi}{8\tilde{G}}}
  = e^{\displaystyle -\frac{\pi}{8cG^p}}\;,
  \lb{SCALCHIRAL}
\end{equation}
to be compared with (\ref{scal}) at $\beta=0$.

\begin{figure}[tbp]
  \begin{center}
    \leavevmode
    \epsfig{file=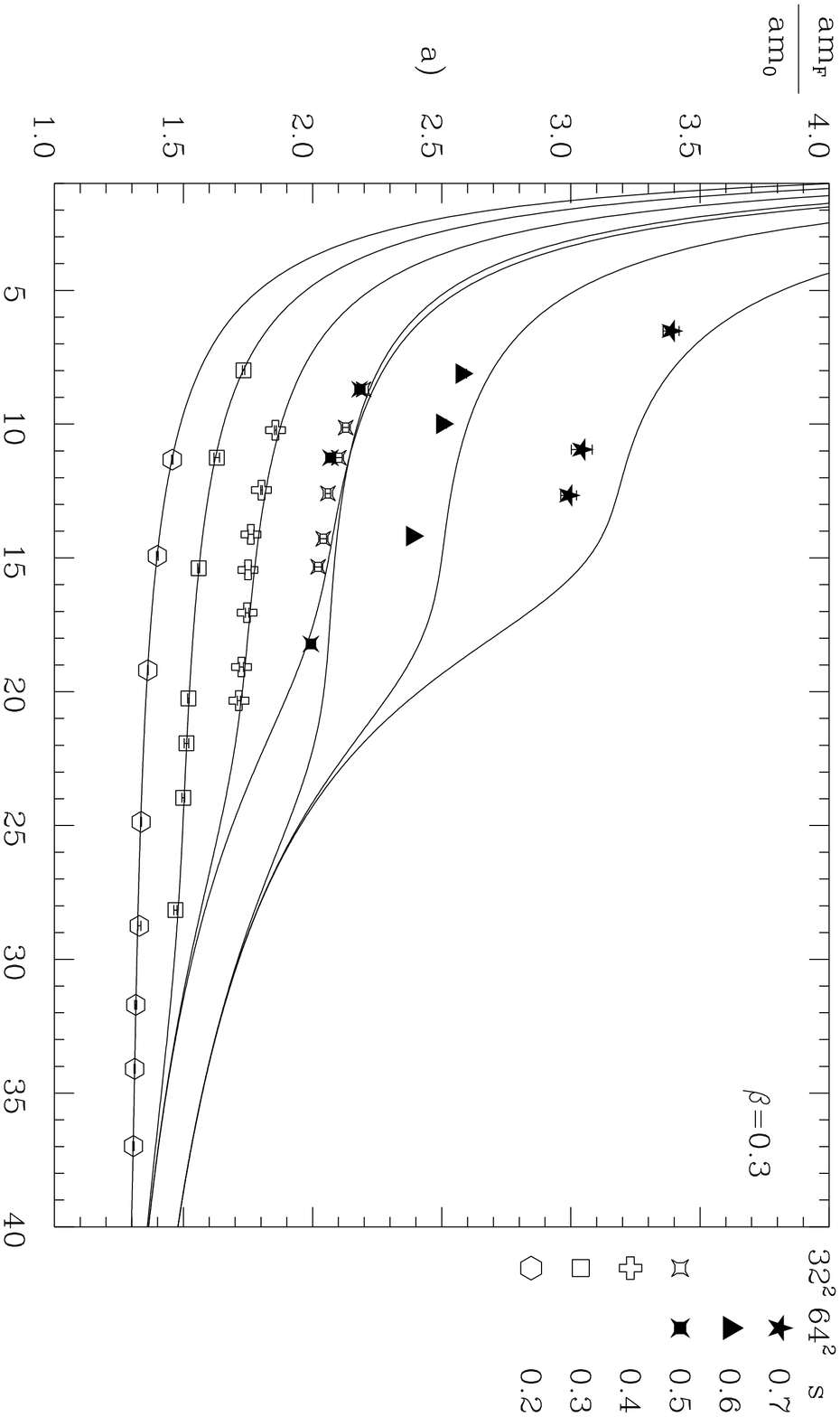,
      width=8.5cm,angle=90}\\
    \vspace*{-1.25cm}
    \epsfig{file=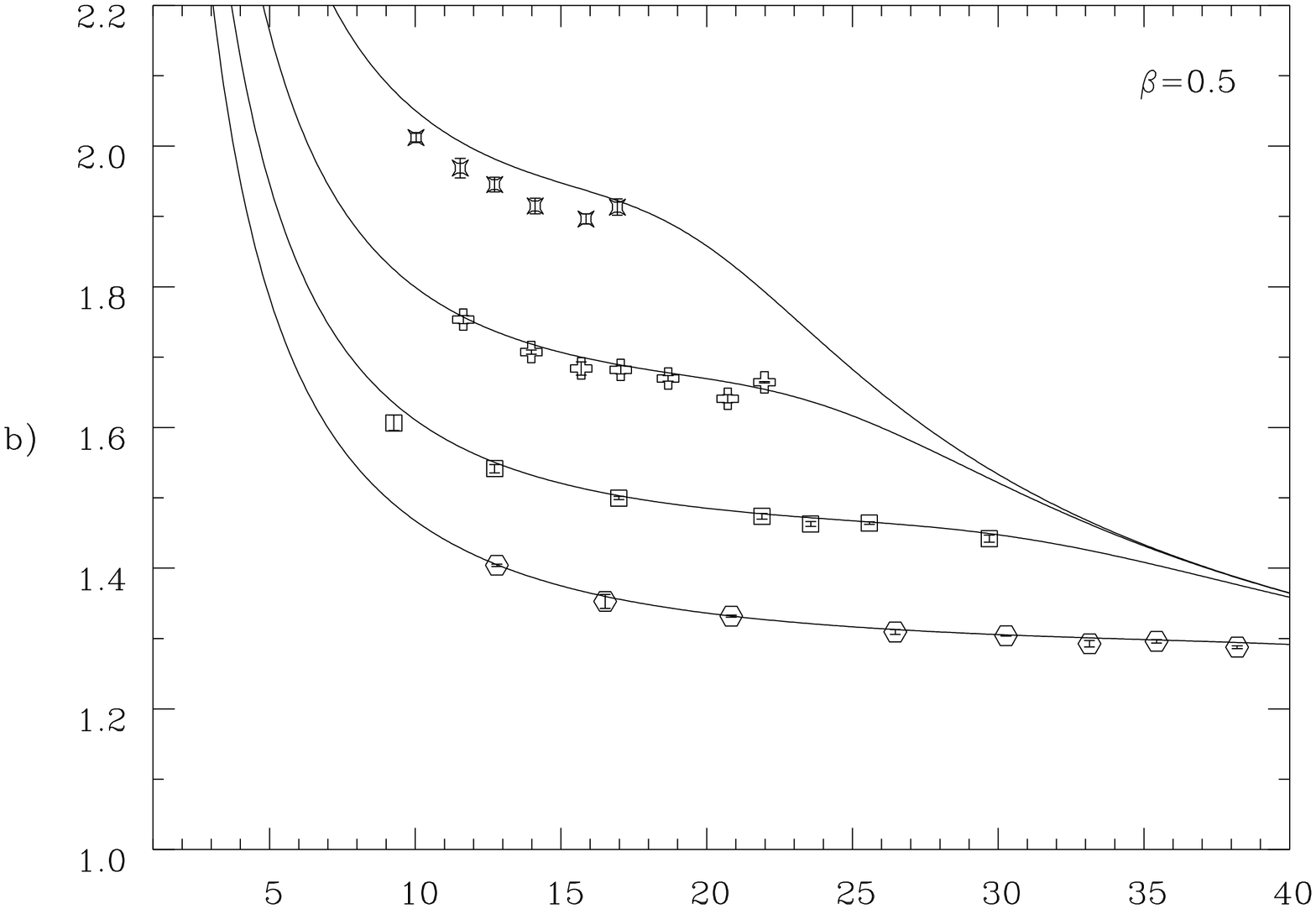,
      width=8.5cm,angle=90}\\
    \vspace*{-1.25cm}
    \epsfig{file=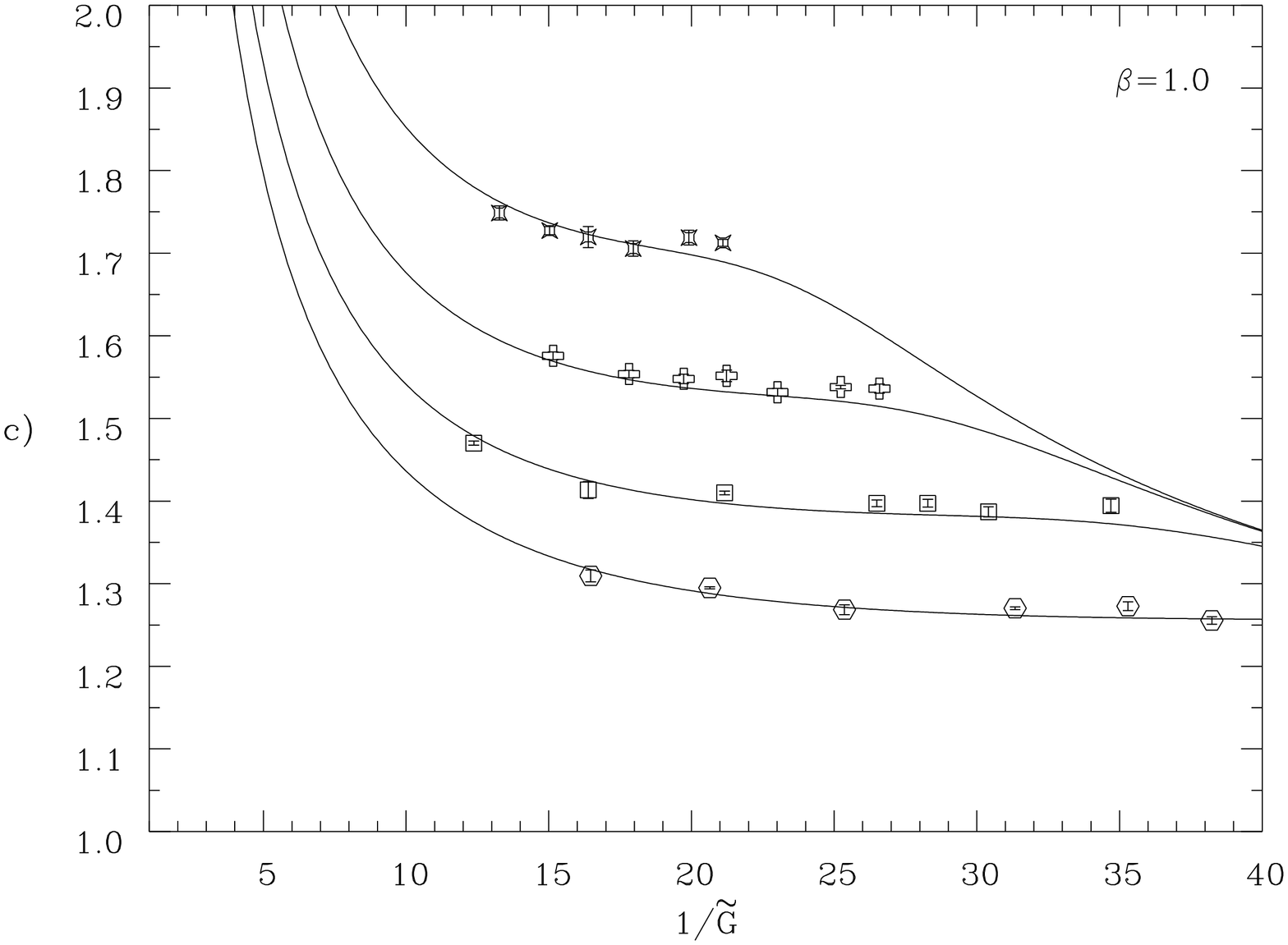,
      width=8.5cm,angle=90}
    %% \vspace{-1.0cm}
    \caption{The ratio $am_F/am_0$ along the lines $s=\mbox{const.}$,
      eq.~(\protect\ref{MBARE}), at $\beta=0.3$, 0.5 and 1.0. The full lines
      are predictions of the SD equations with $\tilde{G}(\beta,G)$ described
      by eq.~(\protect\ref{GG}).}
    \label{fig:8}
  \end{center}
\end{figure}
Of course, our data with rather large $am_0$ and $am_F$ do not show directly
this asymptotic scaling, but only an approach to it, as predicted by the SD
equations. To illustrate this, in Fig.~\ref{fig:8} we compare the data
obtained on lines $s=\mbox{const}$, (\ref{MBARE}), with the predictions for
these lines of the SD equations with $\tilde{G}(\beta,G)$.  As in
Fig.~\ref{fig:5}, we plot the ratio $am_F/am_0$, and the full lines are the
numerical solutions of the SD equations with $\tilde{G}(\beta,G)$. As at
$\beta=0$, Fig.~\ref{fig:5}, we observe a very good agreement for smaller $s$
(larger $am_0$), whereas at larger $s$ at least a qualitative behavior of the
data is reproduced.

The data at $\beta>0$ has been taken on the lines (\ref{MBARE}), which are not
as convenient as the lines (\ref{MBARET}) would be. Nevertheless, one can
obtain the asymptotic behavior on the lines (\ref{MBARE}) as follows:
comparing eqs.~(\ref{MBARE}) and (\ref{MBARET}) we eliminate $am_0$, obtaining
a relation between the pairs of parameters $(G,s)$ and $(\tilde{G},\beta)$,
\begin{equation}  
 (1-s)e^{\displaystyle -\frac{\pi}{8G}s} = 
 (1-\tilde{s})e^{\displaystyle -\frac{\pi}{8\tilde{G}}\tilde{s}}.  \lb{SST}
\end{equation}
Using this relation in eq.~(\ref{SCALINGT}) we obtain for small $G$ (i.e. 
setting $r = F_\mu = 1$)
\begin{equation}
  am_F = \frac{\pi}{2}\frac{1-s}{1-\tilde{s}(\beta,G,s)} 
  e^{\displaystyle -\frac{\pi}{8G}s}\;,
  \lb{SFIT}
\end{equation}
where $\tilde{s}$ is now understood as a function $\tilde{s}(\beta,G,s)$.

\begin{figure}[tbp]
 \begin{center}
  \leavevmode
  \epsfig{file=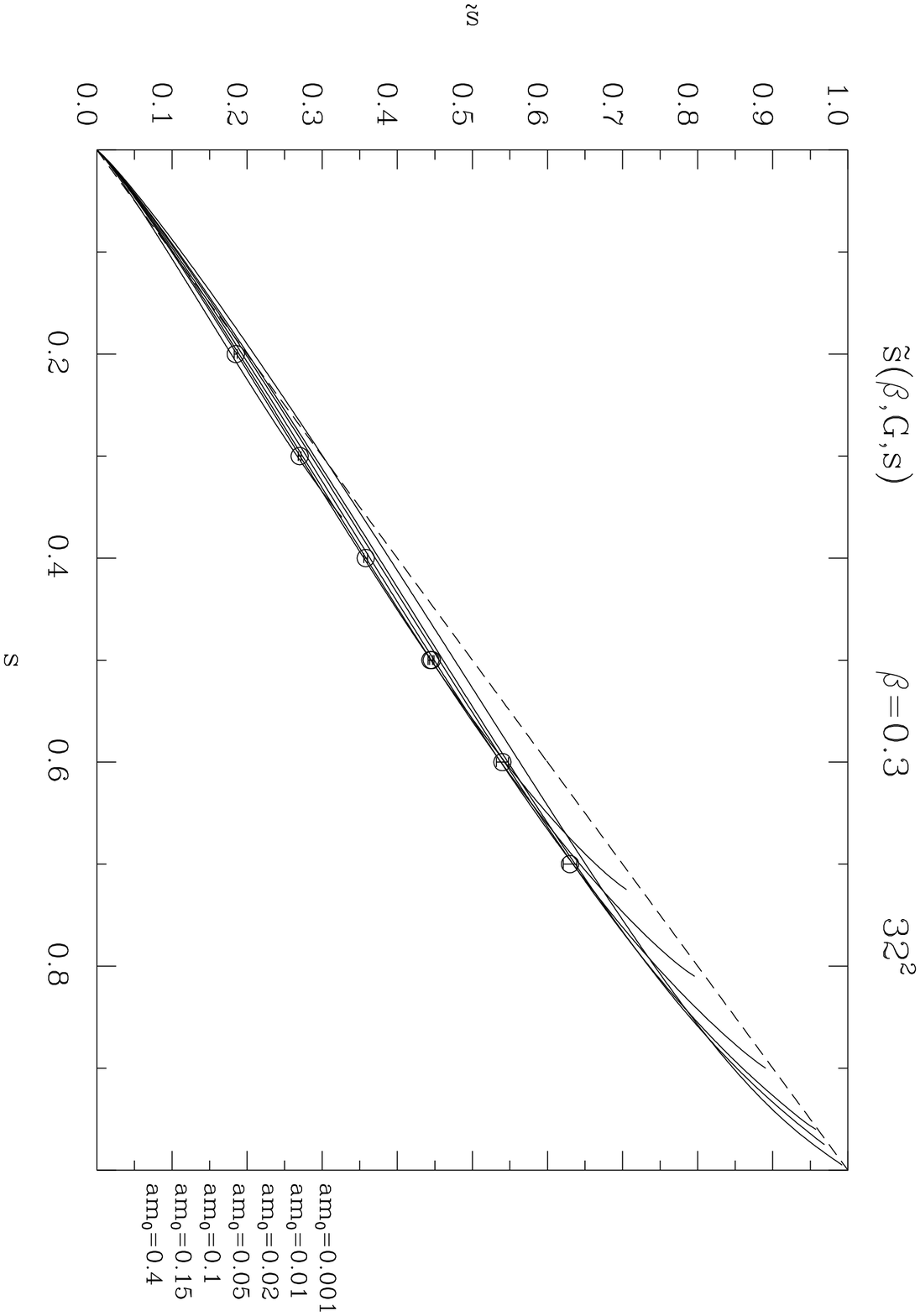,width=10cm,angle=90}
%  \vspace{-1.0cm}
  \caption{The values of $\tilde{s}(\beta,G,s)$ at $\beta=0.3$ for fixed
    $am_0$, determining $G$ by using eq.~(\protect\ref{MBARE}). The curves are
    independent of the lattice volume. The data are from
    Ref.~\protect\cite{FrJe96b}.}
  \label{fig:9}
 \end{center}
\end{figure}

This function $\tilde{s}(\beta,G,s)$ can be determined at each $\beta$ from
the now known $G$-dependence (\ref{GG}) of $\tilde{G}(\beta,G)$. The results
at $\beta = 0.3$ for various fixed values of $am_0$ are shown in
Fig.~\ref{fig:9}.  We observe that, for small $s$, the values of $\tilde{s}$
are nearly independent of $am_0$, and thus of $G$. This explains why the
$\tilde{G}$-dependence of the data in Fig.~\ref{fig:8} looks very similar to
the $G$-dependence of the data at $\beta = 0$, fig.~\ref{fig:5}, up to a --
nearly constant -- factor $(1-s)/(1-\tilde{s})$.  This similarity is easily
observable \cite{FrJe96b}, as no determination of $\tilde{s}(G,s)$ is
required, and thus constitutes a simpler comparison of $am_F$ in the $\chi
U\phi_2$ model at $\beta>0$ with the predictions of the SD equations with
$\tilde{G}$.

The observed agreement between the nonasymptotic data and the SD equations of
the GN$_2$ model with $\tilde{G}$ suggests that also the asymptotic scaling
behavior of the $\chi U\phi_2$ model at $\beta>0$ is described by these
equations, and the model thus belongs to the same universality class. But, of
course, this is only a conjecture, as the region investigated is limited by
the applicability of the truncated SD equations, as well as by the constraints
of a numerical approach.

%%88888888888888888888888888888888888888888888888888888888
\section{Conclusions and discussion}

By introducing an effective four-fermion coupling $\tilde{G}(\beta,G)$ in the
$\chi U\phi_2$ model at $\beta > 0$ we have found that -- within the limits of
a numerical approach -- this model is described by the same SD equations as
the chiral GN$_2$ model with coupling $G$ replaced by $\tilde{G}$. The scaling
behavior, when $\tilde{G} \rightarrow 0$, is thus the same as in the GN$_2$
model in the $G \rightarrow 0$ limit. This is an indication that the $\chi
U\phi_2$ model belongs to the universality class of the GN$_2$ model.

From this we tentatively conclude that the $\chi U\phi_2$ model is
nonperturbatively renormalizable and, though defined on the lattice, possesses
a well defined continuum limit. It is thus an example of a strongly coupled
gauge theory with continuous chiral symmetry, in which the fermion mass is
generated dynamically and the massive fermions are not confined. At least in
two dimensions the shielded gauge mechanism of fermion mass generation,
suggested in Ref.~\cite{FrJe95a}, exists. The $\chi U\phi_2$ model nicely
illustrates this mechanism.

On the other hand, the $\chi U\phi_2$ model is presumably too simple to
provide useful hints how to approach the $d=4$ case.  Our results suggest that
the complexity of the $\chi U\phi_2$ model, and the composite structure of the
fermion $F=\phi^\dagger \chi$, are effects not surviving the continuum limit.
The possibility of describing the scaling properties in terms of the van der
Waals force, represented by the effective four-fermion coupling, shows that in
the renormalized theory the inner structure of the fermion, acquiring its mass
dynamically, is irrelevant.

The study of the renormalizability properties of the strongly coupled $\chi
U\phi_2$ model has been made relatively easy in $d=2$ by its neighborhood to
the well understood GN$_2$ model. Still, the applicability of the SD equations
long before the onset of asymptotic scaling has been a surprise, even in the
GN$_2$ model. The use of these equations has been crucial, since an asymptotic
scaling behavior is evidently not obtainable in numerical simulations, and a
method of extrapolation to the scaling limit, and to the limit of chiral
symmetry, is required.

A plausible explanation why the $\chi U\phi_2$ model is well described by the
SD equations of the GN$_2$ model might be as follows: Integrating out the
gauge and scalar field, one ends up with a pure fermionic theory with many
multi-fermion couplings. These fermions correspond to our fermions $F$, as in
the $\beta=0$ limit. The universality of multi-fermion couplings in $d=2$,
observed e.g. in the studies of the $d=2$ Yukawa model \cite{DeFo93bFoFr94},
suggests that the four-fermion coupling is dominant and sufficient to describe
the data.

Provided that the $\chi U\phi_2$ model at $\beta>0$ really belongs to the
GN$_2$ universality class, the most interesting question left is the
dependence of the effective four-fermion coupling $\tilde{G}$ on $G$ and
$\beta$. We hope that our data for $\tilde{G}(\beta,G)$ will stimulate its
theoretical investigation.

\subsection*{Acknowledgments}
We thank E. Focht, M. G\"ockeler, K.-I. Kondo, M.-P. Lombardo, and P. Rakow
for discussions, and M.\,M. Tsypin for drawing our attention to
Refs.~\cite{CaDa77RaUk78}. The computations have been performed on the
Cray-YMP of HLRZ J\"ulich. The work has been supported by the BMBF and DFG.

% \bibliography{jourabbr,our-papers,gauge,referen,yukawa,cup2}

% end_of_bibliography

\end{document}